\documentclass[preprint,12pt]{elsarticle}

 \usepackage{graphicx, subfigure}
\usepackage{epsfig}
\usepackage{epstopdf}
\usepackage{hyperref}
\usepackage{caption} 
\usepackage{float} 
\usepackage{amssymb}
\usepackage{amsmath}
\usepackage{multirow}

\biboptions{sort}
\biboptions{compress}

\journal{Nuclear Instruments and Methods A}

\begin{document}

\begin{frontmatter}


\title{Study of radioactive impurities in neutron transmutation doped Germanium} 
\author[label1,label2]{S.~Mathimalar}
\author[label1,label2]{N.~Dokania}
\author[label1,label2]{V.~Singh}
\author[label3]{V.~Nanal\corref{cor1}}
\ead{nanal@tifr.res.in}
\cortext[cor1] {Tel.: +91-22-22782333; fax: +91-22-22782133.}
\author[label3]{R.G.~Pillay}
\author[label4]{A.~Shrivastava}
\author[label5]{K.C.~Jagadeesan}
\author[label5]{S.V.~Thakare}
\address[label1]{India-based Neutrino Observatory, Tata Institute of Fundamental Research, Mumbai~400 005, India.}
\address[label2]{Homi Bhabha National Institute, Anushaktinagar, Mumbai 400 094, India.}
\address[label3]{Department of Nuclear and Atomic Physics, Tata Institute of Fundamental Research, Mumbai 400 005, India.}
\address[label4]{Nuclear Physics Divison, Bhabha Atomic Research Centre, Mumbai 400 085, India.}
\address[label5]{Isotope Production and Applicationss Division, Bhabha Atomic Research Centre, Mumbai 400 085, India.}
\begin{abstract}
A program to develop low temperature (mK) sensors with neutron transmutation doped Ge for rare event studies with a cryogenic bolometer has been initiated. For this purpose, semiconductor grade Ge wafers are irradiated with thermal neutron flux from Dhruva reactor at Bhabha Atomic Research Centre (BARC), Mumbai. Spectroscopic studies of irradiated samples have revealed that the environment of the capsule used for irradiating the sample leads to significant levels of~ $^{65}$Zn, $\rm^{110m}$Ag and $^{182}$Ta  impurities, which can be reduced by chemical etching of approximately $\sim50~\mu$m thick surface layer. From measurements of the etched samples in the low background counting setup, activity due to trace impurities of  $^{123}$Sb in bulk Ge is estimated to be $\sim$~1~Bq/g after irradiation.
These estimates indicate that in order to use the NTD Ge sensors for rare event studies, 
a cooldown period of  $\sim$~2 years would be necessary to reduce the radioactive background to $\le$~1~mBq/g. 
\vskip 0.5cm
\end{abstract}
\end{frontmatter}

\section{Introduction}
\label{}
Neutron Transmutation Doped (NTD) Ge thermistors have been widely used as low temperature sensors (in mK range) for bolometric detectors in dark matter searches and neutrino physics~\cite{ntd1,ntd2}. Compared to the conventional metallurgical methods, neutron transmutation doping yields good uniformity and is found to show good reproducibility~\cite{ntd3,ntd4}. The exposure to high neutron dose  can also lead to radioactive contamination of Ge sensors~\cite{cuore} even if starting material is of high  purity.  Such trace radioactivity in sensors can produce significant background for rare event studies like double beta decay. It is therefore important to study and minimize the production of relatively long lived impurities in NTD Ge prior to sensor development. A significant cooldown period for sensors may be needed depending on activity levels~\cite{cuore}.

A program to develop low temperature (mK) sensors with NTD Ge for neutrinoless double beta decay studies with cryogenic bolometer has been initiated. Presently a prototype Tin cryogenic bolometer is under development~\cite{TINTIN}, which will be later housed at the upcoming underground laboratory India-based Neutrino Observatory (INO)~\cite{INO}. Semiconductor grade Ge wafers are irradiated with thermal neutrons from Dhruva reactor at BARC, Mumbai. The Ge samples of varying sizes wrapped in Aluminium were irradiated at designated ports. 
A detailed spectroscopic study of the NTD Ge samples has been carried out in a low background counting setup~\cite{neha} to estimate the radioactive impurities. Chemical etching has been employed to remove the radioactive impurities implanted/diffused close to the surface and an assessment of trace radioactivity in bulk Ge has been carried out.  
An estimate of the cooldown period has been made based on these measurements.

Dependence of radioactive impurities on neutron dose has been investigated. Effect of environment like wrapping material has also been explored within permissible constraints. Section 2 describes experimental details, while Section 3 highlights the radioactive impurity results from spectroscopy  measurements. Conclusions are given in Section 4.

\section{Experimental details}
\subsection{Neutron Irradiation}
Semiconductor grade Ge wafers, supplied by University Wafers, with cleavage planes of $<111>$ (0.4~mm thick, $\rho$ $\sim30~\Omega$~cm) and $<100>$ (1~mm thick, $\rho$  $\ge35~\Omega$~cm) were used in the present studies. The wafers were single side mirror polished. Although the crystal orientation is not expected to have a strong effect on sensor properties, the Ge wafers with $<100>$ orientation are preferred for the ease of cleaving the samples for sensor fabrication. The actual isotopic composition was obtained using Time Of Flight Secondary Ion Mass Spectrometer (TOF-SIMS) measurement~\cite{sims} and is given in  Table~\ref{table1} together with details of n-capture products~\cite{nndc}. Although overall composition is similar, small differences are present in the composition of $<111>$ and $<100>$ oriented wafers, which may depend on the crystal growth condition or the composition of the raw material.  It should be mentioned that both $<100>$ and $<111>$ samples show higher fraction of $^{73}$Ge compared to the natural concentration. This may be a consequence of the starting material being enriched in  $^{73}$Ge, but further details are  not available from the supplier. However, this is not relevant for sensor development since $^{73}$Ge does not contribute to the carrier concentration.  

\begin{table}[H]
\caption{Measured isotopic abundances  of Ge in the wafers used, n-capture reaction and the stable end products are listed together with the decay mode and half-lives of products~\cite{nndc}. The natural concentration of Ge isotopes~\cite{iso} is also included in the column 2 for reference.}
\vspace{0.6em}
 \centering  
  \small\addtolength{\tabcolsep}{-3.5 pt}
\begin{tabular}{ccccl}
\hline
Isotope & \multicolumn{3}{c}{Abundance($\%$)}& \multicolumn{1}{c}{$(n,\gamma)$ reaction product and decay mode (T$_{1/2}$)} \\ 
 & Nat. &$<$111$>$&$<$100$>$ & \multicolumn{1}{c} {} \\ 
\hline
 $^{70}$Ge & 20.6 &21.5(0.2) & 21.9(0.2) & 
{$^{70}\rm Ge\rightarrow ~\rm ^{71}Ge{\xrightarrow {\rm EC(11.4~days)}}~\rm^{71}$Ga }\\
 $^{72}$Ge & 27.4 & 26.8(0.2) & 27.0(0.2) & 
{$\rm^{72}Ge\rightarrow~\rm^{73}Ge$}\\
$^{73}$Ge & 7.8 & 10.8(0.1) & 8.8(0.1) & 
{$\rm^{73}Ge\rightarrow~\rm^{74}Ge$} \\
$^{74}$Ge& 36.5 & 34.1(0.2) &35.1(0.2) & 
$\rm^{74}Ge\rightarrow~\rm^{75}Ge {\xrightarrow {\beta^-(82.8~min)}}~\rm^{75}As$ \\

$^{76}$Ge & 7.7 & 6.8(0.1) &7.2(0.1) & 
$\rm^{76}Ge\rightarrow~\rm^{77}Ge{\xrightarrow {~\beta^-(11.3~hr)~}}~ \rm^{77}As\xrightarrow {\beta^-(38.8~hr)}~\rm^{77}Se $\\
\hline
\end{tabular}
\label{table1}
\end{table}
  
Prior to  irradiation, samples were cleaned in an ultrasonic bath with electronic grade isopropyl alcohol for about 15~min and blow dried with dry N$_2$.   Samples were loaded in a specially designed capsule as per the mandatory procedure for irradiation in the Dhruva reactor.  Different mounting arrangements permissible within operational constraints of the irradiation process in the reactor were tried out to assess the effect of wrapping material and irradiation environment. 
 These consisted of a single sample wrapped in Aluminium foil (Pari, Switzerland, thickness 4.58~$\rm mg/cm^2$), three stacked samples wrapped in a common Aluminium foil and three stacked samples inside a commercial quartz tube. The quartz tube was also wrapped with Aluminium foil.  Both Aluminium and quartz are permissible materials at the Dhruva reactor as the flux attenuation is minimal and there is no resultant long term activity in the wrapping material. In most cases the maximum permissible sample size of 30 mm~$\times$10 mm was used.  The samples were placed inside a cold sealed Aluminium tube capsule (2~cm dia, 4~cm height) for irradiation. In the Dhruva reactor, irradiations are done at fixed positions inside  the reactor core  and the neutron flux is assumed to be uniform due to 4$\pi$ format of neutron emission. The irradiation details like neutron fluences, sample sizes and wrapping materials are given in Table~\ref{table2}. Samples A, B, C are of $<111>$ type, while  D, E, F are of $<100>$ type.  The neutron fluence is estimated from the reactor power, but for a more accurate dose measurement it would be desirable to use a neutron dose monitor (e.g. Fe, Zr, Ni). In the Dhruva reactor, the relative yields of epithermal and fast neutrons (E $>$~0.8 MeV) with respect to thermal neutron yields, are $\sim20\%$ and $\sim1\%$, respectively.

\begin{table}[H]
\centering
\caption{Details of estimated thermal neutron fluence ($\Phi_{\rm th}$) for different Ge samples. The mean irradiation date as well as the duration of irradiation ($t_{irr}$) are also listed in the table.}
\vspace{0.2cm}
\label{table2}
\begin{tabular}{  cccccc }
\hline
\multicolumn{3}{c} {Sample  Details} & Mean date & $t _{irr}$& $\Phi_{\rm th}$ \\
 Label & Wrapping &  Size  & of irradiation& & $\times$~$10^{18}$ \\ 
 &&(mm$^2$)  &(t$_0$)& (days) &(n/cm$^2$)\\
\hline
A &  Al & 10 x 10 &  03/07/2011 & 4.0 & 1.9  \\ 
B & Al & 10 x 10  & 11/08/2012 & 4.1 &14.0 \\ 
C &  Al & 10 x 10  & 11/08/2012 & 4.1 & 9.1  \\ 
D(T1, M1, B1) &quartz & 30 x 10  &  13/09/2013  & 5.7 & 4.6  \\
D(T2, M2, B2) & Al & 30 x 10  &  13/09/2013  & 5.7 & 4.6  \\
E(T3, M3, B3) & quartz & 30 x 10  &  16/11/2013  & 4.1 & 2.1 \\
E(T4, M4, B4) &  Al & 30 x 10  &  16/11/2013   & 4.1 & 2.1  \\
F(T5, M5, B5) & quartz & 30 x 10 &  21/11/2013  & 6.8 & 3.5 \\
\hline
\end{tabular}
\end{table}

After a cooldown period of $\sim$~45 days, individual samples were removed from the irradiation capsule and carefully transferred to separate plastic pouches for spectroscopic measurements. In case of stacked samples, the label M refers to middle sandwiched sample while T  and B refer to the outer samples.  
Some of the larger samples were cut into $\sim$~1 cm $\times$ 1 cm size   pieces after irradiation, which were labeled as L (left), C (center) and R (right).
 Measured activity of the sample with the highest neutron dose was $\sim3~\mu$Sv/hr. 
It should be mentioned that irradiated samples often showed a lack of the lustre and significant improvement was observed after cleaning the NTD Ge samples with HF acid (40~\%). 

\subsection{Radioactive Impurity Measurements}
A specially designed low background  counting setup consisting of $\sim$~70~$\%$ HPGe detector surrounded by low activity Cu~+~Pb  shield \cite{neha} was used for detection of characteristic $\gamma$-rays of radioactive impurities in the irradiated targets.   Data were recorded with a commercial FPGA based 100~MS/s digitizer (CAEN-N6724). Depending on the activity of the sample, counting was done initially at 10~cm from the detector face and later in a close geometry with the sample directly mounted on the detector face. Concentrations of radioactive impurities were obtained from the intensity of the observed $\gamma$-rays after correcting for efficiency, branching ratio and decay during time elapsed since irradiation. The detection efficiency of a photon of given energy is obtained from MC simulations using the detector model as described in Ref.~\cite{neha}. In close geometry,  efficiency corrections due to coincident summing were taken into account.  Typical counting time for each spectrum was about 24 hours.
Spectra for the environmental radioactivity and virgin samples were recorded  for reference. Spectra of the irradiated Aluminium wrapper and the quartz tube were also studied separately. 

\begin{center}
\begin{figure}[H]
 \centering
\includegraphics[scale=0.3]{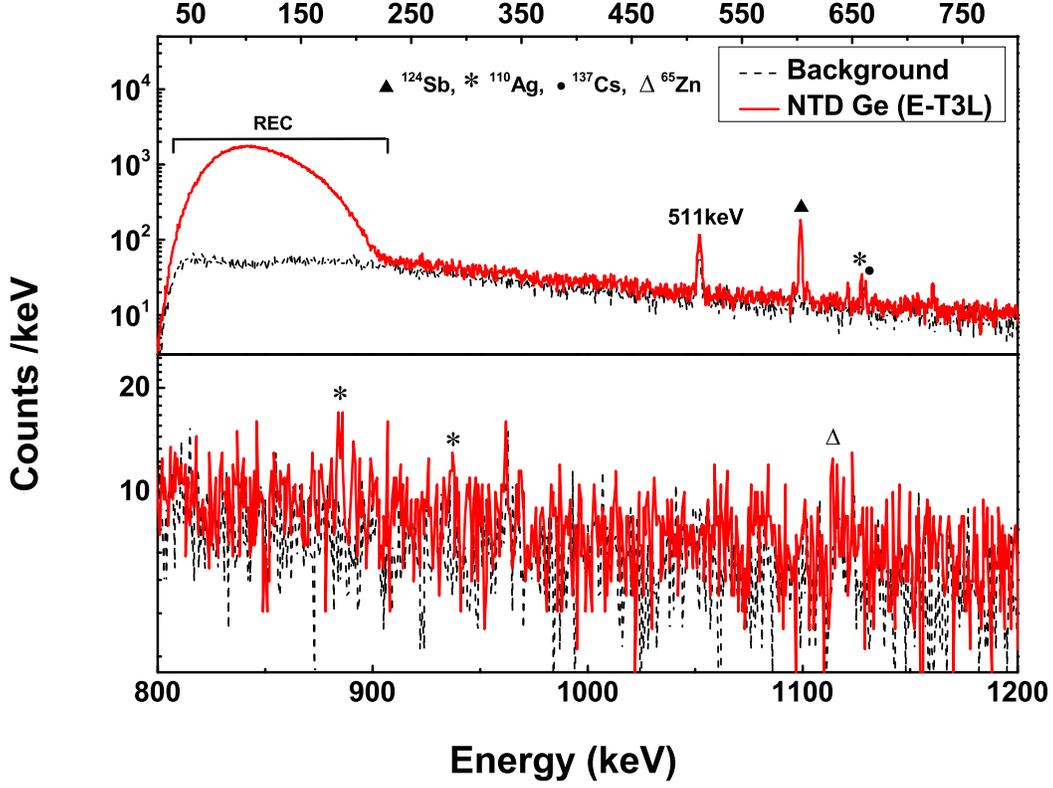}
\caption{(color online) Gamma ray spectra of the NTD Ge  E-T3L sample ($\Phi_{th}$~=~2.1~$\times$~$10^{18}/cm^2$, sample size 10$\times$10 mm$^2$) for  E$_\gamma$~=~0~to~800~keV (top panel) and E$_\gamma$~=~800~to~1200~keV (bottom panel) at t~=~t$_0$~+~125~days (red solid line) together with the environmental radioactivity (black dotted line). Both the spectra are normalized to 12 hours counting time. }
\label{fig1}
\end{figure}
\end{center}

Figure~\ref{fig1} shows $\gamma$-ray spectra of  the irradiated sample E-T3L  (10 $\times$ 10~mm$^2$) after 125 days of cooldown time.  The environmental radioactivity spectrum is also shown for comparison. It should be mentioned that no additional trace impurities were visible in the virgin sample. 
 As can be seen from Table~\ref{table1}, most of the n-capture products of Ge are either stable or have relatively short half-life compared to the initial cooldown period (45 days).  The  $^{71}\rm{Ge}$ has a half-life of 11.4 days and decays mainly by electron capture to the ground state of $^{71}\rm{Ga}$. However, a small fraction that undergoes Radioactive Electron Capture (REC)~\cite{ref5} shows up as a continuous gamma spectrum of $^{71}\rm{Ge}$, which is observed with an end point energy of 225 keV. The gamma rays of $\rm^{110m}$Ag and $^{65}$Zn are also visible in the lower panel. 
 
  Table~\ref{table3} lists the observed radionuclides in various samples together with half-lives, prominent $\gamma$-rays and relative intensities. For unambiguous identification, half-lives of the observed $\gamma$-rays were measured and have been  found to be consistent within 10~$\%$. In addition, wherever applicable, relative intensities of multiple $\gamma$-rays of the given nuclide were also verified. As is evident from the table most of these nuclides are fairly long lived and hence a cause of major concern for low background studies. 
\begin{table}[t]
\centering
\caption{A list of radionuclides and characteristic $\gamma$-rays observed in NTD Ge  samples before etching.}
\label{table3}
\begin{tabular}{cccc}
\hline
Radionuclide& Half-life  & E$_\gamma$  & Relative Intensity \\ 
&&(keV)& ($\%$)\\
\hline
 $^{46}\rm{Sc}$  & 83.79~d & 889.3  & 99.98 \\
 &&1120.5 & 99.99 \\ 
$^{51}\rm{Cr}$ & 27.7~d &  320.1 & 9.91\\ 
 $^{59}\rm{Fe}$ & 44.5~d & 1099.3 & 56.5 \\ 
 &&1291.6  & 43.2\\
 $^{60}\rm{Co}$ & 5.27~y & 1173.2  & 99.85 \\ 
 &&1332.5 & 99.98\\
 $^{65}\rm{Zn}$ & 243.66~d & 1115.5 & 50.04\\ 
 $\rm^{110m}\rm{Ag}$  &    249.76~d      & 657.8  & 95.61 \\ 
 && 884.7 & 75.0\\
 && 937.5 & 35.0\\
  $^{124}\rm{Sb}$      &   60.2~d       & 602.7   & 97.8  \\ 
  && 1691.0 & 47.57 \\
  && 722.8 & 10.76\\
 $^{182}\rm{Ta}$   & 114.74~d & 1121.3   & 35.24 \\
 && 1221.4 & 27.23\\
 && 1231.0 & 11.62\\
 
\hline
\end{tabular}
\end{table}

To investigate the depth dependence of radioactive impurities,  the NTD Ge samples were chemically etched in a controlled manner using H$_2$O$_2$ in  an ultrasonic bath at 80$^\circ$C~\cite{etch} upto a depth of $\sim$ 50 $\mu$m. Samples were cleaned in HF before and after H$_2$O$_2$ etching to remove oxide layers. 
 Typical etching rate observed was 0.3~$\mu$m/min and etched depth was estimated by accurate mass measurement of the sample assuming uniform etching from all sides. 

It should also be mentioned that for n-induced reaction products from wrapping material to get implanted in Ge sample, reactions must take place close to surface of the wrapping material. Hence, the surface trace impurities in wrapping materials were separately studied ($\sim$~few $\mu$m depth) using Energy Dispersive X-ray Analysis (EDAX)~\cite{edax}. 

\section {Results and Discussion}
There are four possible sources of radioactive impurities: 1) neutron induced reaction products of impurities in bulk Germanium,  2) neutron induced reaction products of residual impurities in the Ge surface (resulting from contamination during lapping/polishing/cutting), 3) neutron induced reaction products from wrapping material which can get recoil implanted in Ge and 4) deposition and thermal diffusion of radioactive contaminants from the surrounding environment in the sample capsule resulting from long exposures at high temperatures during irradiation ($\sim80^\circ$C).

\begin{center}
\begin{figure}[H]
 \centering
\includegraphics[scale=0.31]{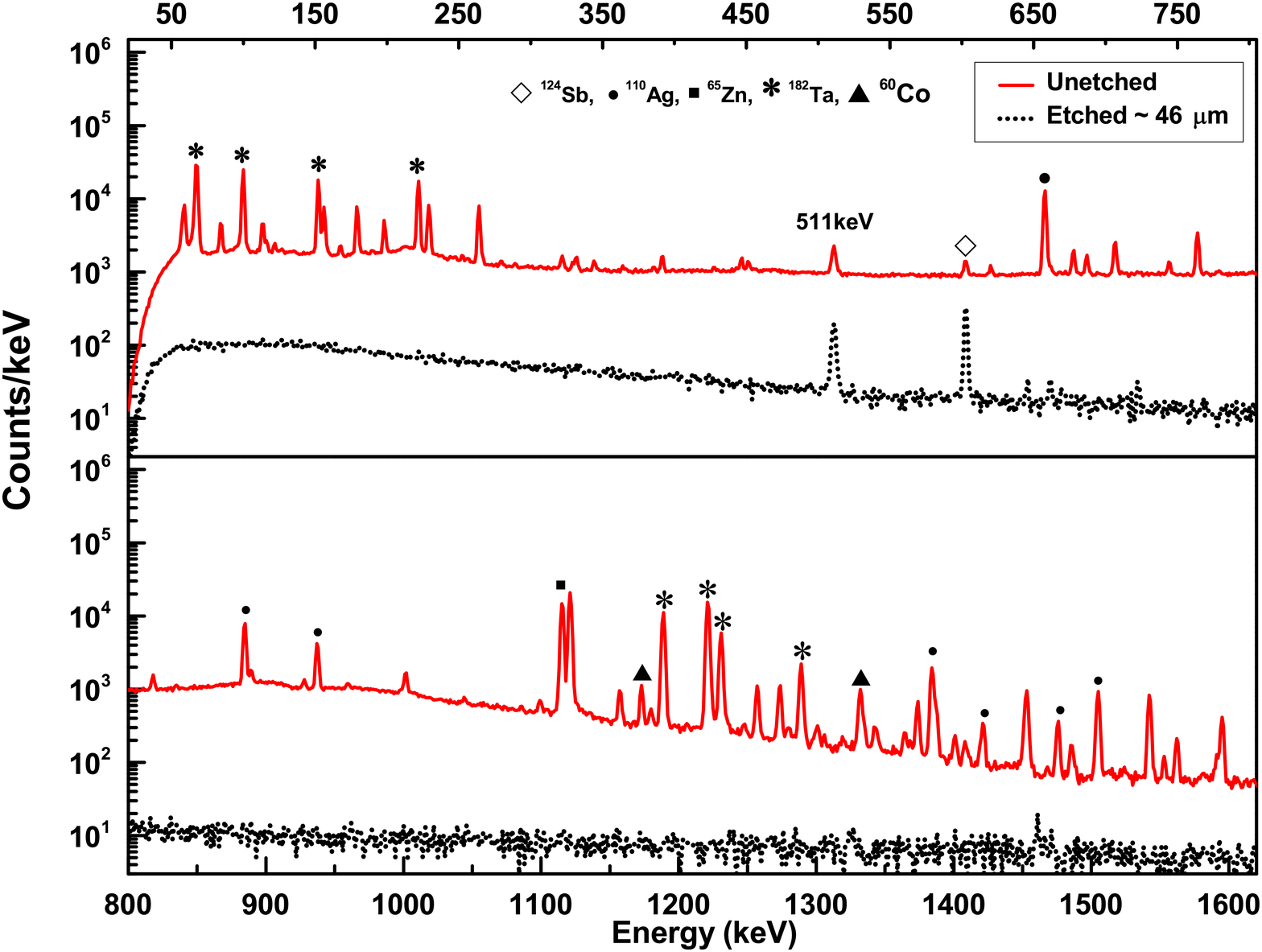}
\caption{(color online) Gamma ray spectra of the NTD Ge sample D-B1  before (red solid line) and after 46~$\mu$m etching (black dotted line) at t~$\sim t_0 + 222$ days for  E$_\gamma$~=~0 to 800 keV (top panel) and E$_\gamma$~=~800~to~1600 keV (bottom panel).  The sample D-B1 (30~$\times$~10~mm$^2$) was mounted in the quartz tube and was exposed to the neuron flunece of  $\Phi_{th}$~=~4.6~$\times$~$10^{18}/cm^2$. Both the spectra are normalized to 12 hours counting time.}
\label{fig2}
\end{figure}
\end{center}

Figure~\ref{fig2} shows $\gamma$-ray spectra of NTD Ge sample D-B1  before and after 46~$\mu$m etching  at t~$\sim t_0 + 222$ days. The sample D-B1 ( {30~$\times$~10~mm$^2$}) was exposed to the highest neutron fluence ($\Phi_{th}$~=~4.6~$\times$~$10^{18}/cm^2$) in the $<100>$ set and clearly shows significantly higher activity as compared to the E samples  (shown in Figure~\ref{fig1}). The REC continuum at low energy is not visible due to larger cooldown time. In the spectrum of the etched sample, it is clearly seen that most of the prominent $\gamma$-rays from the surface impurities are below measurable limits, while $\gamma$-rays from the bulk impurities ($^{124}$Sb) can be seen above the background. 

\begin{figure}[H]
 \centering
\includegraphics[scale=0.4]{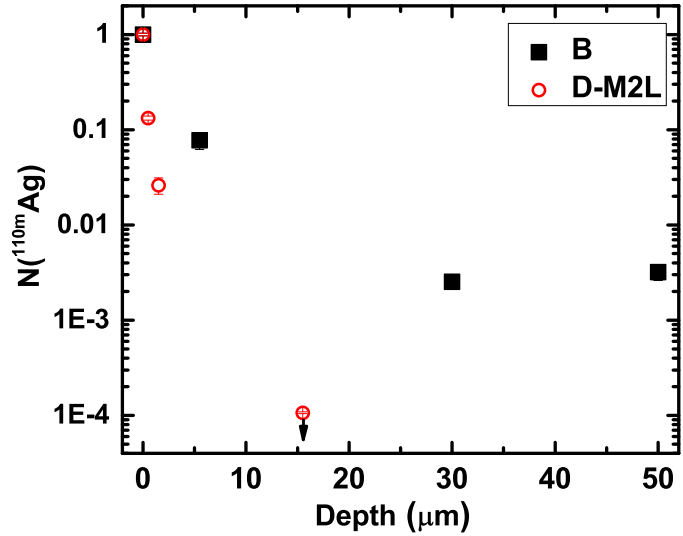}
\caption{(color online) The $\rm^{110m}$Ag activity normalized with respect to the surface value as a function of etching depth for the samples B $<111>$ and D-M2L $<100>$. Errors are smaller than symbol size and arrow indicates the activity below the detection limit.}
\label{fig3}
\end{figure}

 For understanding the origin of radioactive impurities, the observed $\rm^{110m}$Ag activity normalized with respect to the surface value is shown in the Figure~\ref{fig3}  as a function of etching depth for the samples B $<111>$ and D-M2L $<100>$. In both the samples there is a sharp decrease in the activity near the surface, but significant reduction is seen only after about 5 $\mu$m of etching depth.  In case of sample B, $\rm^{110m}$Ag activity was found to remain constant after 10~$\mu$m depth whereas for the sample D-M2L there was no measurable Ag activity after 15~$\mu$m etching. Table~\ref{table4} gives a summary of the observed radioactive impurities at various etching depths (d) in different samples. For most of the samples, the activities of $^{65}$Zn and $^{60}$Co reduced significantly after etching of few~$\mu$m surface layer. In some cases, low level activity became visible after surface level etching (e.g. $^{65}$Zn in E-T3L) due to the improved background. 

\begin{table}[H]
\centering
\caption{Observed radioactive impurities as a function of the etched depth (d) in different samples. When data is not available corresponding to a given etching depth, the cells are left blank.}
\label{table4}
\small\addtolength{\tabcolsep}{-2pt}
\begin{tabular} {llllc }
\hline
Sample& \multicolumn{3}{c}{observed radio-nuclides}& \\
& d~=~0~$\mu$m & d~=~10--20~$\mu$m  &   d~=~50--60$\mu$m \\  \hline
B &   $^{60}$Co,  $^{65}$Zn,  & -& \\
& $\rm^{110m}$Ag, $^{182}$Ta &&$\rm^{110m}$Ag\\ 
C &    $^{60}$Co,  $^{65}$Zn,  & -& \\
& $\rm^{110m}$Ag, $^{182}$Ta &&$^{110m}$Ag\\
D-T1  & $^{60}$Co, $^{65}$Zn,  $\rm^{110m}$Ag,     &  $^{60}$Co, $\rm^{110m}$Ag,    &   $^{124}$Sb \\
& $^{124}$Sb, $^{182}$Ta &  $^{124}$Sb&\\
D-M2L  & $^{60}$Co, $^{65}$Zn,  $\rm^{110m}$Ag,  &  & - \\
& $^{124}$Sb, $^{182}$Ta &  $^{124}$Sb & \\
D-B1  &  $^{60}$Co, $^{65}$Zn,  $\rm^{110m}$Ag, &   $^{124}$Sb, $\rm^{110m}$Ag &   $^{124}$Sb \\
&  $^{124}$Sb, $^{182}$Ta  & &\\
E-T3L&   $^{124}$Sb, $\rm^{110m}$Ag & $^{124}$Sb, $^{65}$Zn & $^{124}$Sb\\
F-B5 & $^{124}$Sb, $\rm^{110m}$Ag, $^{65}$Zn & $^{124}$Sb&  $^{124}$Sb\\
  \hline
\end{tabular}
\end{table}
 Since implanted impurities will be restricted to sub-micron layers near the surface, the observation of activities  upto 10~$\mu$m layer is surprising and not easy to understand. This could result from thermal diffusion of impurities which will depend on both temperature and time (Ref.~\cite{agdiff}). Given the time scale of neutron irradiation ($\sim$ 4 to 5 days) at 80$^o$C there might be a  possibility of heavy nuclei like Ag thermally diffusing at micrometer scale. The surface contamination during the wafer processing stage is also a possibility.
 It should be mentioned that even in case of D samples, which had highest neutron fluence amongst $<100>$ set ($\Phi_{\rm th}\sim 4.6$ $\times$ $10^{18}/{\rm cm}^2$),  some traces of $\rm^{110m}$Ag and $^{182}$Ta could be seen till 40~$\mu$m depth. 
 No measurable activity was found in the NTD A sample, which had a nearly 3 years of cooldown time. However, since a very small size sample ($\sim$~15~mg) was used for spectroscopic studies, no limits on radiopurity of the sample were extracted.
Samples B, C from  $<111>$   and  D, E, F from $<100>$  showed different bulk impurities. The only measurable radioactivity present in $<100>$ Ge after $\sim50~\mu$m etching was  $^{124}$Sb. While the $^{124}$Sb activity was not seen in $<111>$ Ge samples, they showed activities of  $^{65}$Zn  (222~$\pm$~87 mBq/g) and $\rm^{110m}$Ag (225~$\pm$~67 mBq/g) even after 50~$\mu$m etching and $\sim$~1.6 years of cooldown period. 

 Table~\ref{table5} lists the measured activities for various NTD Ge samples, 150~days after the irradiation. 
 The etched samples from $<100>$ set did not show any measurable $\rm^{110m}$Ag activity and only detection limits are mentioned. In Ref.~\cite{cuore}, which reports the sensor development for CUORE experiment, it is mentioned that the residual activity of  $\rm \ge 1 mBq/g$ after 3 years is of significant concern. Hence in the present work, we have estimated the required cooldown time to achieve a similar level of activity, which is listed in the last column.  It is evident that for $<100>$ samples (D/E/F), with the expected dose $\Phi_{\rm th}\sim 1-5$ $\times$ $10^{18}/{\rm cm}^2$, approximately 2 years of cooldown period after irradiation is essential. 
The higher radioactivity levels of $\rm^{110m}$Ag in samples B and C indicate that material from these samples will be unsuitable for sensors in low counting experiment. Further, since the neutron fluence of samples B and C are higher than $5\times10^{18}$ n/cm$^{2}$ (as given in Table 2), these samples will be near metallic and are unsuitable for sensor fabrication. It should be mentioned that the commercial NTD Ge sensor (AdSem, Inc~\cite{adsem}) showed much higher levels of $^{65}$Zn and $\rm^{110m}$Ag, possibly due to other materials used in contact fabrication.

\begin{table}[H]
\centering
\caption{Measured radioactivity and estimated cooldown period (T$_{\rm cool}$) for reduction of the radio-activity below $< 1$~mBq/g for various NTD Ge samples. Detection limits are indicated, wherever the activity could not be measured above the background.}

\label{table5}
\begin{tabular} {ccccc }
\hline
Sample& \multicolumn{2}{c}{Activity~(mBq/g)}&T$_{\rm cool}$ \\
&\multicolumn{2}{c}{(t$_0$ + 150 days)}& (yr)\\
& $\rm^{110m}$Ag & $\rm^{124}$Sb &  \\  \hline
B &  3018(465)  & $<$~26.6 &9 \\
C & 743(222)    & $<$~18.6& 7  \\
D-B1  & $<$~2.0   & 420(9)& 1.9  \\
E-T3&  $<$~2.5* &201(20)& 1.7\\
F-B5 &   $<$~2.9 & 344(12)& 1.8\\
  \hline
    \multicolumn{4}{l} {* Estimated from sample E-T3L} 
\end{tabular}
\end{table}

As mentioned earlier, Alessandrello $\it{et~al.}$~\cite {cuore} have also measured the residual radioactivity in NTD Ge thermistors for a similar neutron fluence, name\-ly, $\Phi_{\rm th}~\sim~3.36$~$\times$~$10^{18}$/cm$^2$. They have reported several isotopes like $^{75}$Se, $^{74}$As and $^{68}$Ge resulting from fast neutron induced reactions during irradiation. 
Elliott $\it{et~al.}$~\cite{gerda} have also reported the formation of isotopes like $^{65}\rm{Zn}$, $^{54}\rm{Mn}$  and $^{60}\rm{Co}$ in high energy neutron induced reactions.  
In the present case though $^{65}\rm{Zn}$ activity was seen at surface of the samples, the $\gamma$-rays corresponding to fast neutron induced reaction products (namely, $^{75}$Se, $^{74}$As, $^{68}$Ge, $^{54}\rm{Mn}$, $^{60}\rm{Co}$ and  $^{65}\rm{Zn}$)   are not visible in the irradiated-etched samples.  
The detection limits on activities of $^{75}$Se, $^{74}$As, $^{68}$Ge for the sample F-B5 at t$\sim t_0$+150 days are given in Table~\ref{table6} together with the values reported in Ref.~\cite{cuore}, measured immediately after the irradiation. While the limit on $^{68}$Ge activity is lower in the present case, the $^{74}$As activity is similar to that in Ref.~\cite{cuore}. The detection limit for the $^{75}$Se activity in the present case is significantly  worse than the measured value of Ref.~\cite{cuore}. It should be noted that present measurements are carried out after a cooldown time of $\sim$~150 days, where the short lived activities like $^{74}$As would have significantly decayed. Also, the yield of (n,$\alpha$) or (n,3n), (n,p) reactions responsible for $^{75}$Se and $^{68}$Ge  production will depend on high energy neutron flux, which can be different in two cases.

\begin{table}[H]
\centering
\caption{Estimated limits on fast neutron induced reaction products in Ge for sample F-B5 after a cooldown period of 150 days. Data from Ref.~\cite{cuore} immediately after irradiation is also shown for comparison.}

\label{table6}
\small\addtolength{\tabcolsep}{-3pt}
\begin{tabular} {clcccc}
\hline
Radionuclide & Possible reactions& T$_{1/2}$ & Present   & Data from \\
&&& detection limit&Ref~\cite{cuore}\\
& & days & mBq/g & mBq/g\\
\hline
 $^{68}$Ge  & $^{70}$Ge(n,3n), $^{72}$Ge(n,5n) & 270 & $<$51 & 150(10) \\

$^{75}$Se & $^{72}$Ge($\alpha$,n), $^{73}$Ge($\alpha$,2n),   & 119.78 & $<$15.2& 3.3(0.8)\\ 
& $^{74}$Ge($\alpha$,3n)&&\\

$^{74}$As & $^{74}$Ge(p,n), $^{73}$Ge(n,$\gamma$)  & 17.72 & $<$2.8 & 4400(750)\\

 $^{60}$Co & $^{59}$Co(n,$\gamma$) & 1925.28 & $<$1.9 & - \\

  \hline
\end{tabular}
\end{table}

The observed residual activity of $^{124}$Sb results predominantly from the $^{123}$Sb(n,$\gamma$) reaction with thermal neutrons. The contribution from the fast neutrons can be neglected since the flux for E$_n>$~1~MeV is smaller by a factor  $\sim$~100 and the cross-section for n-capture is smaller by a factor of $\sim$~50.  The concentration of the reaction product is related to that of the parent isotope ($N_{impurity}$) by the following relation:

\begin{equation}\label{eqn1}
{N^{product}_\gamma} = \frac{N_{impurity}\times{\sigma_{c}} \times{\phi_{\rm th}}\times(1-e^{-\lambda t_{irr}})}{\lambda}
\end{equation}
where ${\sigma_{c}}$ is the thermal neutron capture cross-section (to the ground state and/or excited state as the case may be)~\cite{ref3}, $\lambda$ is the decay constant, $t_{irr}$ is the duration of the irradiation and
${\phi_{\rm th}}$ is the thermal neutron flux expressed in units of $\rm neutrons.cm^{-2}.s^{-1}$. The ${\phi_{\rm th}}$ is assumed to be uniform during the irradiation period, i.e. ${\phi_{\rm th}}$ = ${\Phi_{\rm th}}$/t$_{irr}$.

 The ${N^{product}_\gamma~}$ is computed from the measured intensity of $\gamma$-ray (N$_\gamma$) during the counting time interval of $t_1$ to  $t_2$  (measured with respect to end of the irradiation) and is given by
 
\begin{equation}\label{eqn2}
{N^{product}_\gamma~} ={\frac{ N_{\gamma}}{\epsilon_{\gamma}\times I_{\gamma}\times(e^{-\lambda t_{1}}-e^{-\lambda t_{2}})} }
\end{equation}
where $\epsilon_{\gamma}$ and $I_{\gamma}$ are the photo-peak detection efficiency and branching ratio of the $\gamma$-ray, respectively. 
 
\begin{figure}[h]
 \centering
\includegraphics[scale=0.4]{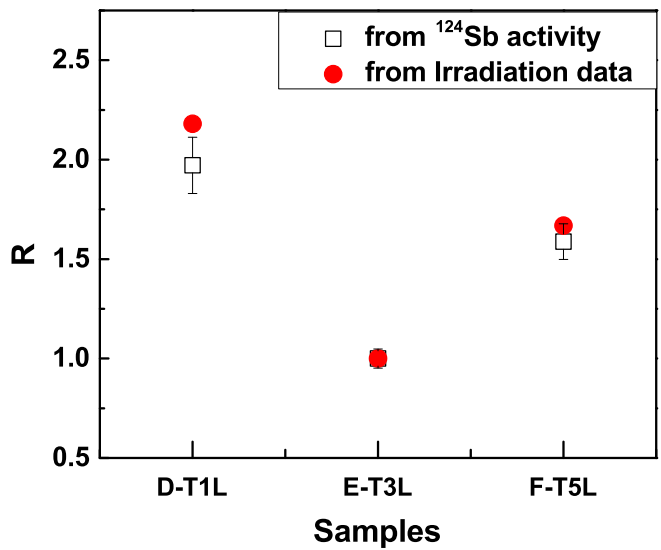}
\caption{(color online) Relative neutron fluence (R) of samples D and F  with respect to the sample E  from the residual $^{124}$Sb activity (open square) and from the irradiation data (filled circle).  }
\label{fig4}
\end{figure}

\begin{table}[h]
\centering
\caption{ Estimated trace $^{123}\rm{Sb}$ impurities from the residual radioactivity of $^{124}\rm{Sb}$ in etched  $<100>$ NTD Ge samples. The etched depth for each sample is indicated in the bracket. Systematic errors in detector efficiency~(about 5~$\%$)~\cite{neha} are not included and only statistical errors are shown.}
\label{table7}
\begin{tabular}{cccccc}
\hline
Parent  & E$_\gamma^{\rm product}$  &  \multicolumn{3}{c}{Concentration (ppt)} \\
  isotope & (keV)  & D-B1(46~$\mu$m) & E-T3L(42~$\mu$m) & F-B5 (52~$\mu$m)\\ \hline
$^{123}\rm{Sb}$ & 602.7 &  115(2) & 119(12) & 123(4)\\
\hline
\end{tabular}
\end{table}
 Table~\ref{table7} lists the estimated bulk impurity concentration of $^{123}\rm{Sb}$ from the observed activity of $^{124}\rm{Sb}$ using equations 1 and 2. The isotope $^{121}\rm{Sb}$, which is more abundant, produces short-lived activity ($\sim$~2.7~days) and could not be observed in the present studies. However, considering the presence of $^{123}\rm{Sb}$ in Ge as indicative of $\rm^{nat}\rm{Sb}$  and  assuming the natural abundances for $^{121}\rm{Sb}$ and $^{123}\rm{Sb}$ (57$\%$ and 43$\%$, respectively~\cite{iso}), the bulk impurity of $\rm^{nat}\rm{Sb}$ in $<100>$ Ge is estimated to be 277(14) ppt.
 Figure~\ref{fig4} shows a plot of relative neutron fluence (R) of sample D and F  with respect to the E sample extracted from  $^{124}$Sb activity (open square) and from the irradiation data (filled circle). The good agreement between these two indicate that the  observed bulk impurity concentration of  $^{123}$Sb is similar in different samples of $<100>$. It may be noted that the observed bulk impurity level of $^{123}$Sb $\sim 100~$ppt is much below the sensitivity of SIMS (ppm) and hence could not be measured in the SIMS. Futher,  Alessandrello et al.~\cite{cuore} have reported much smaller bulk impurity concentration of  $^{123}$Sb in Ge, namely, $<$~1~ppt. The $^{123}$Sb impurity needs to be minimized as it leads to high energy gamma ray background ($E_\gamma > 2$~MeV) and therefore it would be desirable to use detector grade Ge as a starting material instead of the device grade.

 The EDAX analysis of  Aluminium wrapper and the quartz tube had shown a purity level of $\sim$~99\%. 
The irradiated Aluminium wrapper showed very high levels of $\rm^{110m}$Ag, $^{65}$Zn and $^{182}$Ta, while the quartz tube showed several additional $\gamma$-rays of $^{54}$Mn, $^{58}$Co and $^{134}$Cs. Any reaction products originating from (n, $\gamma$), (n, $\alpha$), (n, p) reactions on $^{27}$Al result in stable/short-lived products such as $^{28}$Al (T$_{1/2}$ = 2.2~m), $^{24}$Na (T$_{1/2}$ = 15~h) and $^{27}$Mg(T$_{1/2}$ = 9.5~m), respectively. Hence these short-lived products could not be observed after $\sim$~45 days.
 For a neutron fluence of $\Phi_{\rm th}\sim$ 4.6 $\times$ 10$^{18}/\rm cm^2$,  the measured $\rm^{110m}$Ag activity  in the Aluminium wrapper and the quartz was  $\sim$~2.2 kBq/g  (corresponding to $\sim$~0.2~ppm of $^{109}$Ag in $\rm^{nat}$Al) and $\sim$~0.07 kBq/g  (corresponding to $\sim$~0.02~ppm of $^{109}$Ag in SiO$_2$), respectively,  after a cool down period of $\sim$~150~days.  The surface activity of $\rm^{110m}$Ag in the corresponding Ge samples, namely, D-T2 (wrapped in the Aluminum) and D-T1 (in the quartz tube) was 56(2)~Bq/g and 7.2(0.4)~Bq/g, respectively.  It is evident that the Ge samples wrapped in Aluminium showed higher surface activity as compared to those in the quartz tube.   
 
Further improvements like irradiation in a sealed quartz capsule to reduce effect of environment and pre-etching of samples to remove residual surface impurity prior to irradiation are under consideration.  It should be mentioned that the performance of the NTD Ge sensor made from the sample D was tested and found to be satisfactory. Initial  results at $\sim$~100 mK have been reported in the Ref.~\cite{mathi-wolte}.

\section{Conclusions}

The development of  low temperature (mK) sensors with neutron transmutation doped Ge for rare event studies with a cryogenic bolometer has been initiated. For this purpose, semiconductor grade Ge wafers were irradiated with thermal neutrons at the Dhruva reactor at BARC, Mumbai.
Irradiated Ge samples have been studied in the low background counting setup and all peaks in the $\gamma$-spectra were identified.
Chemical etching of the surface removed most of the long lived impurities,  indicating that  these impurities can arise either from residual surface contamination of the sample or due to diffusion during irradiation from  the sample capsule environment. To reduce activity from residual surface contamination, pre-etching of samples prior to n-irradiation will be highly  desirable.
The chemical etching does not affect the performance of the sensor, but is desirable even from thermometry consideration since radioactive impurities can act as a standing heat load at mK temperature. 
   For the desired neutron fluence of $1-5$ $\times$ 10$^{18}$/cm$^2$, removal of 50~$\mu$m surface layer is found to be adequate for this purpose. The samples loaded in the quartz tube are found to have lower radioactivity than those wrapped in Aluminium. 
 The observed radioactive impurities ~$\sim$~1~Bq/g in the  bulk Ge, estimated after chemical etching of the samples, implies that a cooldown period of  $\sim$~2 years would be necessary before sensors made from these samples can be used in rare decay studies requiring ultra low background ($\le $~1~mBq/g).

\section{Acknowledgements}
We thank Ms. S. Mishra and Mr. K.S.~Divekar for help in sample preparation, Mr.~M.S. Pose and Mr. K.V. Anoop for assistance in $\gamma$-ray counting setup, Mr. R.D. Bapat and Ms. B.A. Chalke for help during EDAX measurement, Mr.~K.G.~Bhushan for performing SIMS measurement and Dr. V.M.~Datar for the useful discussions.


\begin{thebibliography}{00}
\bibliographystyle{elsarticle-num}
\bibitem{ntd1} O. Martineau \textit{et al.,} Nuclear Instruments and Methods in Physics Research Section A {\bf530} (2004) 426. 
\bibitem{ntd2}C. Arnaboldi \textit{et al.,} (CUORE Collaboration), Nuclear Instruments and Methods in Physics Research Section A {\bf518} (2004) 775. 
\bibitem{ntd3} E.E.~Haller, Infrared Phys. $\bf{25}$ (1985) 257.
\bibitem{ntd4}K.M.~ltoh, E.E.~Haller, W.L.~Hansen, J.W.~Beeman, J.W.~Farmer, A.~Rudnev, A.~Tikhomirov and V.I.~Ozhogin, Appl. Phys. Lett. {\bf64}~(1994) 2121.
\bibitem{cuore}A. Alessandrello, C. Brofferio, D.V. Camin, O. Cremonesi, E. Fiorini, A. Giuliani, M. Pavan, G. Pessina, E. Previtali, L. Zanotti, Nuclear Instruments and Methods in Physics Research Section B $\bf{93}$ (1994) 322.
\bibitem{TINTIN}  V. Nanal, EPJ Web of Conferences $\bf{66}$ (2014) 08005.
\bibitem{INO}N.K.~Mondal, Pramana $\bf{79}$ (2012) 1003.
\bibitem{neha} N. Dokania, V. Singh, S. Mathimalar, V. Nanal, S. Pal, R.G. Pillay, Nuclear Instruments and Methods in Physics Research Section A {\bf745} (2014) 119.
\bibitem{sims}Peter Williams, Ann. Rev. Mater. Sci. {\bf15}~(1985) 517.
\bibitem{nndc} http://www.nndc.bnl.gov.
\bibitem{iso} M. Berglund, M.E. Wieser, Isotopic compositions of the elements 2009 (IUPAC Technical Report), Pure Appl. Chem. {\bf 83} (2011) 397.
\bibitem {ref5}B.L. Saraf, J. Varma, C.E. Mandeville, Physical Review {\bf91} (1953) 1216.
\bibitem{etch} N. Cerniglia and P. Wang, Journal of The Electrochemical Society {\bf109} (1962) 508.
\bibitem{edax}John C. Russ, Fundamental of Energy Dispersive X-ray Analysis, Butterworth-Heinemann Ltd (1984).
\bibitem{agdiff} Ling Y. Wei, J. Phys. Chem. Solids, {\bf18} (1961) 162.
\bibitem {adsem}http://adsem.com/index.html.
\bibitem{gerda}S.R.~Elliott, V.E.~Guiseppe, and B.H.~LaRoque, R.A.~Johnson, S.G.~Mashnik, Physical Review C {\bf82} (2010) 054610.
\bibitem{ref3}R.B. Firestone, V.S. Shirley, Table Of Isotopes, {\bf 1}, 8th Edition.
\bibitem{mathi-wolte}S.~Mathimalar, V.~Singh, N.~Dokania, V.~Nanal, R.~G.~Pillay, S. Pal, S.~Ramakrishnan, A. Shrivastava, Priya Maheshwari, P.K.~Pujari, S.~Ojha, D.~Kanjilal, K.C.~Jagadeesan, S.V. Thakare, IEEE WOLTE conference, (2014) 10.1109/WOLTE.2014.6881014.

\end{thebibliography}
\end{document}